\documentclass[aps,prb,reprint,superscriptaddress]{revtex4-1}

\usepackage{graphicx}
\usepackage{bm}

\begin{document}

%\preprint{}

\title{Effects of chiral helimagnets on vortex states in a superconductor}


\author{Saoto Fukui}
\author{Masaru Kato}
%\thanks{}
\affiliation{Department of Mathematical Sciences, Osaka Prefecture University, 1-1, Gakuencho, Sakai, Osaka 599-8531, Japan}
\author{Yoshihiko Togawa}
\affiliation{Department of Physics and Electronics, Osaka Prefecture University, 1-1, Gakuencho, Sakai, Osaka 599-8531,Japan}


\date{\today}

\begin{abstract}
We have investigated vortex states in chiral helimagnet/superconductor bilayer systems under an applied external magnetic field $H_{\rm appl}$, using the Ginzburg-Landau equations.
Effect of the chiral helimagnet on the superconductor is taken as a magnetic field $H_{\rm CHM}$, which is perpendicular to the superconductor and oscillates spatially.
For $H_{{\rm appl}}=0$ and weak $H_{{\rm CHM}}$, there appear pairs of up- and down-vortices. %which are parallel and antiparallel to $H_{\rm appl}$, respectively.
Increasing $H_{\rm appl}$, down-vortices gradually disappear, and number of up-vortices increases in the large magnetic field region.
Then, up-vortices form parallel, triangular, or square structures.
\end{abstract}

%\keywords{}

\maketitle


%\begin{eqnarray}
% \alpha \left| \psi \right|^2 + \beta \left| \psi \right|^2 \psi + \frac{1}{4m} \left\{ \frac{\hbar}{i} \nabla - \frac{2e}{c} \bm{A} \right\}^2 \psi = 0 \label{gl_1} \\
% \mbox{\boldmath $J$} = \frac{e\hbar}{2mi} \left( \psi^\ast \nabla \psi - \psi \nabla \psi^\ast \right) - \frac{2e^2}{mc} \psi^\ast \psi \bm{A} \label{gl_2}
%\end{eqnarray}

%\begin{eqnarray}
% {\rm curl}~(\mbox{\boldmath $H$} - \mbox{\boldmath $H$}_{{\rm ext}}) = \frac{4\pi}{c} \mbox{\boldmath $J$} \label{max_1} \\
% {\rm curl}~\bm{A} = \mbox{\boldmath $H$} \label{vector_potential}
%\end{eqnarray}

%\begin{eqnarray}
% \frac{ \left(H_{{\rm ext}}\right)_z}{\Phi_0/\xi_0^2}_z (x) &=& \frac{H_0}{\Phi_0/\xi_0^2} {\rm cos} \Biggl( 2 {\rm sin}^{-1} \left({\rm sn} \left( \frac{\sqrt{\beta}}{k}x | k \right) \right) \Biggr. \nonumber \\
%                                                            &=& \Biggl. + \pi \Biggr) + \frac{H_{{\rm appl}}}{\Phi_0/\xi_0^2}  \label{external_field}
%\end{eqnarray}
%\begin{equation}
% \beta = \frac{2\mu_B H_{{\rm appl}}}{\sqrt{J^2 + D^2}S} \label{beta-h}
%\end{equation}

\section{Introduction \label{Introduction}} 
Vortices are key features for type-II superconductors in determining a critical magnetic field and a critical current.
%In general, Abrikosov lattices are formed in the homogeneous magnetic field\cite{Abrikosov}.
In general, when a homogeneous magnetic field is applied to a superconductor,
vortices form a triangular lattice called the Abrikosov lattice\cite{Thinkham, de_Gennes, Abrikosov}.
Under an external current, vortices may move and then electric resistivity appears.
Also, configurations of vortices affect motions of vortices.
So, controlling vortex states is important for applications of superconductivity.

Recently, ferromagnet (FM)/superconductor (SC)  systems have been studied\cite{F/S_Review,F/S_dot,F/S_bilayer}.
A magnetic structure of a FM causes a magnetic field in the superconductor and this magnetic field affects the superconductivity.
It was found that the vortices appear and because of a pinning effect on vortices, transport properties such as a critical current of the SC changes. 
%For example, in the regular array of square magnetic dots with out-of-plane magnetization on the SC, the magnetic field from dots is applied to the SC\cite{F/S_dot}.
%The magnetic field between neighboring dots is antiparallel to the direction of the magnetization of the dots.
%In this situation, the critical magnetic field in this system becomes larger than that in only the SC because the larger magnetic field must be applied to cancel the opposite magnetic field.
%In another work, the SC/FM layer system is considered\cite{F/S_bilayer}.
%Due to the magnetic structure of the FM, vortices are formed spontaneously.
%In addition, when the FM has a stripe domain structure, vortices which correspond to the magnetic domain are formed.
%This vortex structure restricts the movement of vortices.

The FM is the most effective magnetic material on the SC.
However, there are other magnetic materials that have a large effect on the vortex state.

\begin{figure}[b]
 \includegraphics[width=8.5cm]{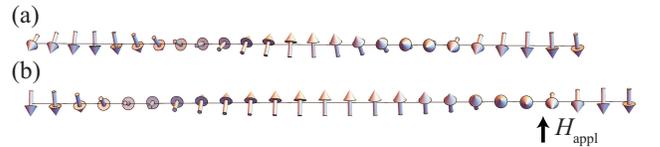}
 \caption{Schematics of magnetic structures of (a) a chiral helimagnet at zero magnetic field and (b) a chiral soliton lattice under an applied magnetic field $H_{\rm appl}$.
          These spins form a helical rotation along some direction.}
 \label{Fig1}
\end{figure}

Recently, a chiral helimagnet (CHM) has been studied actively in the field of magnetism\cite{Togawa_CSL, Togawa_MR, kishine_spin_current, Togawa_twist}.
The chiral helimagnet consists of spins that form a helical rotation along some direction in Fig.\ref{Fig1} (a).
This helical spin ordering comes from a competition between two interactions; the Dzyaloshinsky-Moriya interaction and the ferromagnetic exchange interaction.
The Dzyaloshinsky-Moriya interaction is expressed as $-\bm{D} \cdot (\bm{S}_1 \times \bm{S}_2)$, where $\bm{S}_1$ and $\bm{S}_2$ are localized nearest neighbor spins along one direction\cite{Dzyaloshinsky, DM_Moriya}.
$\bm{D}$ is the Dzyaloshinsky-Moriya interaction vector (DM vector).
This interaction causes directions of $\bm{S}_1$ and $\bm{S}_2$ are perpendicular.
The direction of this vector is determined by a crystal structure in the CHM.
On the other hand, the ferromagnetic exchange interaction is expressed as $-J \bm{S}_1 \cdot \bm{S}_2$, where $J$ is an exchange coefficient ($J>0$).
This interaction causes all spins are parallel.
When $|\bm{D}|$ is much smaller than $J$, directions of $\bm{S}_1$ and $\bm{S}_2$ are slightly deviated from the ferromagnetic structure because of the competition between two interactions.
This deviation leads to the helically clockwise rotated structure of spins along the direction of the vector $\bm{D}$.
%The DM vector determines the direction of the helical rotation.

Under an applied magnetic field, the magnetic structure of the CHM transforms into an incommensuate magnetic structure, which is called a chiral soliton lattice (CSL) in Fig.\ref{Fig1} (b).
%In the chiral soliton lattice, there are some solitons and we can regard solitons as the $360^\circ$ domain wall.
The period of solitons can be controlled by the magnetic field.
In the experiment, magnetic structures of the CHM and the CSL are observed using Lorentz microscopy analysis and small-angle electron diffraction\cite{Togawa_CSL}.
This formation of CSL causes peculiar properties\cite{Togawa_MR}.
%The chiral soliton lattice has forced ferromagnetic domains periodically devided by 360$^\circ$ domain walls.
The CHM is expected be used in controlling of the spin current in the field of the spintronics\cite{kishine_spin_current} and a novel magnetic processor\cite{Togawa_twist}.
%In recent work, the superconductivity of the chiral helimagnet Cr$_{1/3}$NbS$_2$ at the very low temperature was reported\cite{Togawa_JPS}.
%Cr$_{1/3}$NbS$_2$ is a layer structure of 2H-NbS$_2$ intercalated by Cr.
%The space group is $P6_322$.
%2H-NbS$_2$ is a SC at very low teperature $T_c=6.05\pm0.2$K\cite{NbS_2}.
%The superconductivity of Cr$_{1/3}$NbS$_2$ is considered to come from the formation of the chiral helimagnet/SC bilayer system like Fig.\ref{bilayer}.

We expect that the CHM affects the SC strongly.
These influences may be unlike to those from the FM.

In this paper, we investigate effects of the CHM on the SC in a CHM / SC bilayer system, using the Ginzburg-Landau (GL) equations.
In particular, we focus on the effect on vortex states in the SC.
In Sec.~\ref{Methods}, we introduce a model and numerical methods.
In Sec.~\ref{Results}, we show results about distributions of order parameters (A) and vortex states in the CHM / SC bilayer system (B, C).
In Sec.~\ref{Summary}, we summarize this paper.
In Appendix, we discribe some coefficients, which are used in the GL equations in Sec.~II.

\section{Methods \label{Methods}}
\begin{figure}[t]
 \includegraphics[width=7cm]{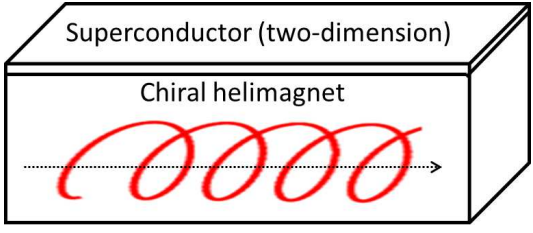}
 \caption{The CHM / SC bilayer system}
 \label{Fig2}
\end{figure}

We consider a CHM / SC bilayer system in Fig.\ref{Fig2}.
The effect of the CHM on the SC is taken as an external magnetic field $\bm{H}_{\rm CHM}$.
In this study, the SC layer is much thinner than the CHM layer.
So, this SC layer is considerd as a two-dimentional system.
In addition, the CHM layer is thick and $\bm{H}_{\rm CHM}$ is assumed to have only the perpendicular component to the interface between the CHM and the SC.
On the other hand, effects of the SC on the CHM (for example, the Meissner effect) are neglected. 
Then, we investigate vortex states in the two-dimensional superconducting systems under $\bm{H}_{\rm CHM}$\cite{ISS2014}.

To obtain vortex states in the two-dimensional superconducting systems under $H_{\rm CHM}$, we solve the GL equations\cite{d-dot}.
Under a magnetic field, the GL free energy $G(\psi,\bm{A})$ is written as,
\begin{eqnarray}
 G(\psi, \bm{A}) &=& \int_\Omega \left( f_n + \alpha \left| \psi \right|^2 + \frac{\beta}{2} \left| \psi \right|^4 \right) d\Omega \nonumber \\
                 & & + \int_\Omega \left\{ \frac{1}{2m_s} \left| \left( -i\hbar \nabla - \frac{e_s \bm{A}}{c} \right) \psi \right|^2 \right. \nonumber \\
                 & & \left. + \frac{\left| \bm{h} \right|^2}{8\pi} - \frac{\bm{h} \cdot \bm{H}}{4\pi} \right\} d\Omega \label{gl_free},
\end{eqnarray}
where $\psi$ is a superconducting order parameter and $f_n$ is the Helmholtz free energy of the normal state.
$\alpha$ is a coefficient which depends on temperature $T$ as $\alpha(T) = \alpha'(T - T_c)$, where $\alpha'$ is the constant ($\alpha' > 0$) and $T_c$ is a critical temperature of the SC.
$\beta$ is a positive constant, $m_s$ is a effective mass of the SC, and $e_s$ is a effective charge of electrons in the SC.
$\bm{H}$ is an external field, and $\bm{h} = \nabla \times \bm{A}$ is the local magnetic field, where $\bm{A}$ is a magnetic vector potential.
We use an alternative form,
\begin{equation}
 \epsilon(\psi, \bm{A}) = G + \int_\Omega \left\{ \frac{\alpha^2}{2\beta} + \frac{\bm{H} \cdot \bm{H}}{8\pi} - f_n + \frac{\left( {\rm div}\bm{A} \right)^2}{8\pi} \right\} d\Omega. \label{alternative}
\end{equation}
We add $({\rm div}\bm{A})^2$ term to Eq.(\ref{alternative}) to insure the Coulomb gauge ${\rm div}\bm{A}=0$.
For minimizing $\epsilon$ with respect $\psi$, $\psi^\ast$, and $\bm{A}$, we obtain following equations,
\begin{eqnarray}
 & & \int_\Omega \left[ \left( i\nabla \tilde{\psi} - \tilde{\bm{A}} \tilde{\psi} \right) \left( -i\nabla \tilde{\psi}^\ast - \tilde{\bm{A}} \left( \delta\tilde{\psi} \right)^\ast \right) \right. \nonumber \\
 & & \left. + \left( i\nabla \left( \delta\tilde{\psi} \right) - \tilde{\bm{A}} \left( \delta\tilde{\psi} \right) \right) \left( -i \nabla \tilde{\psi}^\ast - \tilde{\bm{A}} \tilde{\psi}^\ast \right) \right. \nonumber \\
 & & \left. + \frac{1}{\xi^2} \left( \left| \tilde{\psi} \right|^2 - 1 \right) \left( \tilde{\psi} \left( \delta\tilde{\psi} \right)^\ast + \tilde{\psi}^\ast \left( \delta\tilde{\psi} \right) \right) \right] d\Omega = 0, \label{gl_1} \\ \nonumber \\ \nonumber \\
 & & \int_\Omega \left[ \kappa^2 \xi^2 \left\{ {\rm div} \tilde{\bm{A}} \cdot {\rm div} \left( \delta\tilde{\bm{A}} \right) + \nabla \times \tilde{\bm{A}} \cdot \nabla \times \left( \delta\tilde{\bm{A}} \right) \right\} \right. \nonumber \\ 
 & & \left.+ \left| \tilde{\psi} \right|^2 \tilde{\bm{A}} \cdot \left( \delta\tilde{\bm{A}} \right) - \frac{i}{2} \left\{ \tilde{\psi}^\ast \left( \nabla \tilde{\psi} \right) - \tilde{\psi} \left( \nabla \tilde{\psi}^\ast \right) \right\} \tilde{\bm{A}} \right] d\Omega \nonumber \\
 & & = \kappa^2 \xi^2 \int_\Omega \frac{2\pi}{\Phi_0} \bm{H} \cdot \nabla \times \left( \delta\tilde{\bm{A}} \right) d\Omega, \label{gl_2}
\end{eqnarray}
where $\delta\tilde{\psi}, \delta\tilde{\bm{A}}$ are variations of the order parameter and the vector potential.
$\tilde{\psi}$ and $\tilde{\bm{A}}$ are normalized order parameter and vector potential as,
\begin{eqnarray}
 \tilde{\psi} &=& \frac{\psi}{\sqrt{|\alpha|/\beta}} \\
 \tilde{\bm{A}} &=& \frac{2\pi}{\Phi_0}\bm{A}.
\end{eqnarray}
$\Phi_0$ is the quantum flux, $\Phi_0 = ch/2e$, where $e$ is an electron, $e_s=2e$.
$\xi$ is the coherence length, which depends on the temperature, $\xi^2(T) = \hbar^2/(4\pi \left| \alpha(T) \right|)$.
$\kappa$ is the GL parameter, $\kappa = \lambda/\xi$, where $\lambda$ is the penetration length.
In this study, we use following boundary conditions: 
$\tilde{\bm{j}} \cdot \bm{n} = 0$, where $\bm{n}$ is the normal vector to the surface, and $\left( i\nabla \tilde{\psi} - \tilde{\bm{A}} \tilde{\psi} \right) \cdot \bm{n} = 0$ at the boundary.
%In addition, edges in this system are free.

To solve these two Eqs.(\ref{gl_1}) and (\ref{gl_2}), we use the two-dimentional finite element method.
$\tilde{\psi}$ and $\tilde{\bm{A}}$ are expanded using area coordinates $N_j^e$ ($j = 1$, $2$, and $3$) for $e$-th element,
\begin{eqnarray}
  \tilde{\psi}{(\bm{x})}   &=& \sum_e \sum_{j=1}^{3} N_j^e(\bm{x}) \tilde{\psi_j^e} \label{linear_psi}\\
  \tilde{\bm{A}}{(\bm{x})} &=& \sum_e \sum_{j=1}^{3} N_j^e(\bm{x}) \tilde{\bm{A}_j^e}, \label{linear_a}
\end{eqnarray}
where $\tilde{\psi_j^e}$ and $\tilde{\bm{A}_j^e}$ are values of order parameter and vector potential at $j$-th node in $e$-th element.
We set test functions $\delta \tilde{\psi}$ and $\delta \tilde{\bm{A}}$ as,
\begin{eqnarray}
 \delta \tilde{\psi} &=& N_i^e(\bm{x})~~~~(i=1,2,3) \\
 \delta \tilde{\bm{A}} &=& N_i^e(\bm{x})e_\gamma ~~~~(i=1,2,3,~\gamma=x,y,z) 
\end{eqnarray}
Then, Eqs.(\ref{gl_1}) and (\ref{gl_2}) become as,
\begin{eqnarray}
 & & \sum_j \left[ P_{ij}^e (\{ \tilde{\bm{A}} \}, \{ \tilde{\psi} \}) + P_{ij}^{e2R} (\{ \tilde{\psi} \}) \right] {\rm Re}~\tilde{\psi_j^e} \nonumber \\
 & & + \sum_j \left[ Q_{ij}^e (\{ \tilde{\bm{A}} \}) + Q_{ij}^{e2} (\{ \tilde{\psi} \}) \right] {\rm Im}~ \tilde{\psi_j^e} = V_i^{eR} (\{ \tilde{\psi} \}),  \label{gl_3} \\
 & & \sum_j \left[ -Q_{ij}^e (\{ \tilde{\bm{A}} \}) + Q_{ij}^{e2} (\{ \tilde{\psi} \}) \right]  {\rm Re}~\tilde{\psi_j^e} \nonumber \\
 & & + \sum_j \left[ P_{ij}^e (\{ \tilde{\bm{A}} \}, \{ \tilde{\psi} \}) + P_{ij}^{e2I} (\{ \tilde{\psi} \}) \right] {\rm Im}~ \tilde{\psi_j^e} = V_i^{eI} (\{ \tilde{\psi} \}), \nonumber \\ \label{gl_4} \\
 & & \sum_j R_{ij}^e(\{ \tilde{\psi} \}) \tilde{A_{jx}^e} + \sum_j S_{ij}^e \tilde{A_{jy}^e} = T_i^{ex} - U_i^{ey}, \label{gl_5} \\
 & & -\sum_j S_{ij}^e \tilde{A_{jx}^e} + \sum_j R_{ij}^2 (\{\tilde{\psi}\}) \tilde{A_{jy}^e} = T_i^{ey} + U_i^{ex}. \label{gl_6}
\end{eqnarray} 
Coeffficients are given in Appendix.

\begin{figure}[h]
 \includegraphics[width=8cm]{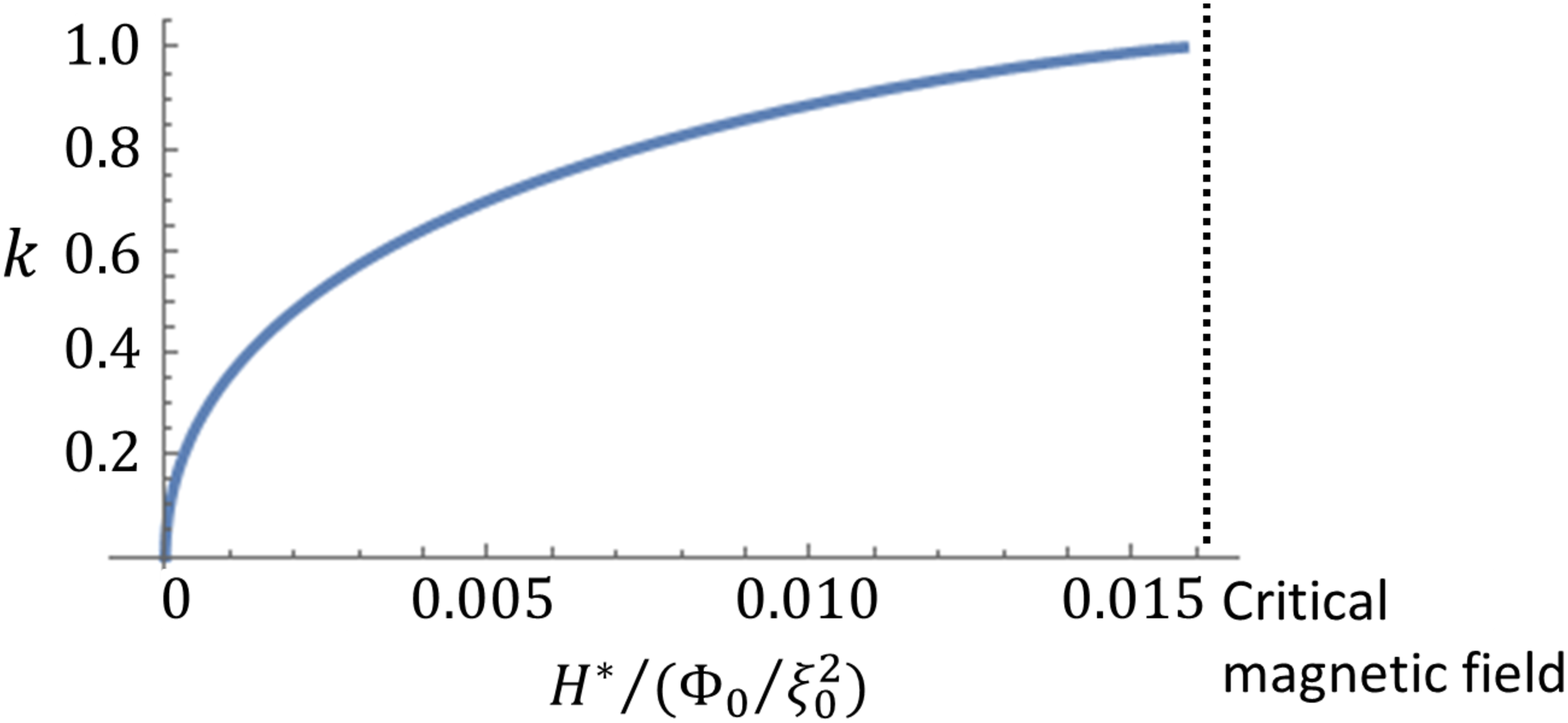}
 \caption{The relation between the modulus of the Jacobi's elliptic function and the applied magnetic field for $D/J = 0.16$}
 \label{Fig3}
\end{figure}

The magnetic field from the CHM, $\bm{H}_{\rm CHM}$ is included in the external magnetic field $\bm{H}$ in Eq.(\ref{gl_2}).
%In addition, $U_i^x$ and $U_i^y$ are functions of the external magnetic field $H$, so the effect of the CHM is also taken in these terms.
$\bm{H}_{\rm CHM}$ is obtained from the Hamiltonian,
\begin{eqnarray}
 \mathcal{H} &=& -2J \sum_n \bm{S}_n \cdot \bm{S}_{n+1} + \bm{D} \cdot \sum_n \bm{S}_n \times \bm{S}_{n+1} \nonumber \\
             & & - 2\mu_B H_{{\rm appl}} \sum_n S_n^z, \label{hamiltonian_chiral}
\end{eqnarray}
where $\bm{S}_n$ is the spin of $n$-th site.
The first term is a ferromagnetic exchange interaction term, where $J$ is the magnitudes of the exchange coefficient.
The second term is the Dzyaloshinsky-Moriya interaction term, where $\bm{D}$ is the DM vector.
The third term is the Zeeman energy term, where $H_{\rm appl}$ is the homogeneous applied magnetic field.
$\mu_B$ is the Bohr magneton.
%Each termas are the ferromagnetic exchange interaction, Dzyaloshinsky-Moriya interaction, and the Zeeman energy.
We represent $\bm{S}_n$ in terms of the polar coordinates as,
\begin{equation}
 \bm{S}_n = S(\sin{\theta}_n \cos{\varphi}, \sin{\theta_n} \sin{\varphi}, \cos{\theta}_n), \label{spin}
\end{equation}
and minimize the energy with respect to $\theta_n$\cite{kishine}.
We obtain the Sine-Gordon equation,
\begin{equation}
 \frac{d^2\theta}{dx^2} - H^\ast \sin{\theta} = 0, \label{Sine-Gordon}
\end{equation}
where $H^\ast$ is a normalized applied magnetic field,
\begin{equation}
 H^\ast = \frac{2\mu_BH_{{\rm appl}}}{\xi_0^2S^2\sqrt{J^2 + D^2}}. \label{H^ast}
\end{equation}
We assume $\xi_0 = a$, where $a$ and $\xi_0$ are a lattice constant and a coherence length for the SC at $T=0$, respectively.
The solution is 
\begin{equation}
 \theta(x) = 2\sin^{-1}{\left[ {\rm sn} \left( \frac{\sqrt{H^\ast}}{k}x \right) \right]} + \pi, \label{sine_sol}
\end{equation} 
where $k$ is the modulus of the Jacobi's elliptic function sn$(x|k)$ and is determined by,
\begin{equation}
 \frac{\pi \phi}{4\sqrt{H^\ast}} = \frac{E(k)}{k}, \label{k_det}
\end{equation}
where $\phi = \tan^{-1}{\left( D / J \right)}$ and $E(k)$ is the complete elliptic integral of the second kind.
In Eq.(\ref{k_det}), when the applied magnetic field $H^\ast$ increases, the modulus $k$ also increases monotonically from $0$ to $1$, which is shown in Fig.\ref{Fig3}.

\begin{figure}[h]
 \includegraphics[width=7cm]{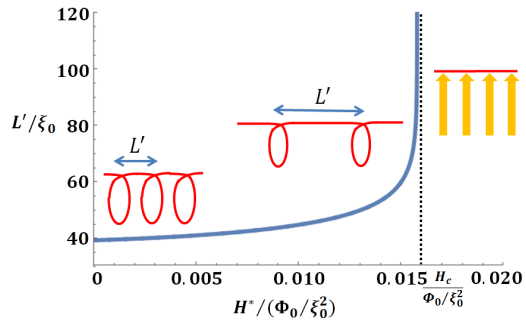}
 \caption{The relation between the helical period and the applied magnetic field for $D/J = 0.16$.}
 \label{Fig4}
\end{figure}

The period of the CHM $L'$ is given as,
\begin{equation}
 \frac{L'}{\xi_0} = \frac{2kK(k)}{\sqrt{H^\ast}}, \label{L_period}
\end{equation}
where $2K(k)$ is the period of the ${\rm sn}(u|k)$ function and $K(k)$ is the complete elliptic integral of the first kind.
This relation (Eq.(\ref{L_period})) is shown in the Fig.~\ref{Fig4}.
Increasing the applied magnetic field, the period becomes longer and spins form the CSL.
The period increases rapidly before transition to the ferromagnetic.

Using the solution of the Sine-Gordon equation in Eq.(\ref{sine_sol}) to the GL equations, we can obtain vortex states in the magnetic field from the CHM.

\section{\label{Results} Results and Discussions}
We solve Ginzburg-Landau equations and study distributions of the order parameter and vortex states.
We take the Ginzburg-Landau parameter $\kappa = \lambda/\xi = 10,$ the temperature $T = 0.3T_c$.
The ratio of the strength of the Dzyaloshinsky-Moriya interaction to that of the ferromagnetic exchange interaction is taken from Cr$_{1/3}$NbS$_2$ as $D/J=0.16$\cite{D/J}.
The external magnetic field $\bm{H}_{\rm ext}$ is given by the sum of the magnetic field from the CHM $\bm{H}_{\rm CHM}$ and the applied magnetic field $\bm{H}_{\rm appl}$.
We only consider $z$-component of the external magnetic field $\bm{H}_{\rm ext}$.
Using Eq.(\ref{sine_sol}), it is given by,
\begin{eqnarray}
 \left(\bm{H}_{{\rm ext}}\right)_z (x) &=& H_0 {\rm cos} \left\{ 2 {\rm sin}^{-1} \left[{\rm sn} \left( \frac{\sqrt{H^\ast}}{k}x | k \right) \right] + \pi \right\} \nonumber \\
                                                     & & +H_{{\rm appl}}.  \label{external_field}
\end{eqnarray}
%where $\Phi_0$ is a quantum flux.
%This magnetic field is normalized by the $\Phi_0/\xi_0^2$.
%The first term is the magnetic field from the CHM and the second term is the external applied field.
%In the first term, the factor cos$\theta(x)~( \theta(x)$ is given in Eq.(\ref{sine_sol}) ) represents a distribution of spins, which is used as the magnetic field directly.
Here, we assume that the distribution of the magnetic field from the CHM (the first term in Eq.(\ref{external_field})) is proportional to the distribution of spins in the CHM (Eq.(\ref{spin})).
The factor $H_0$ represents the magnitude of the magnetic field from the CHM. %, which corresponds to the strength of the CHM.
In the following, the system size is set as $7.0 L' \times 20 \xi_0$, where $L'$ is the period of the CHM in Eq.(\ref{L_period}).
When $H_{{\rm appl}}/(\Phi_0/\xi_0^2)=0.00$, $L'/\xi_0$ becomes approximately $39.2699$.%, where we normalize $H_{\rm appl}$ and $L'$.

\subsection{Effects of the CHM on distributions of the order parameter}
\begin{figure}[t]
\centering
  \includegraphics[scale=0.25]{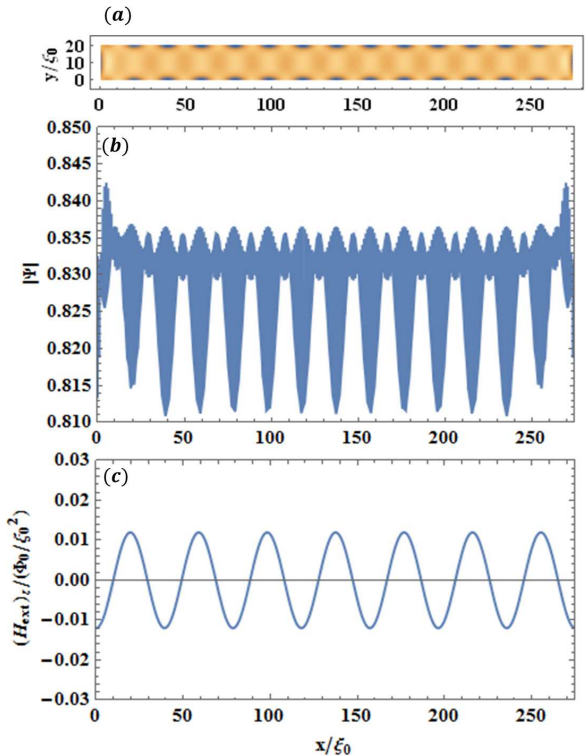}
  \caption{(a) The distribution of the order parameter. (b) The amplitude of the order parameter. (c) The distribution of the magnetic field from the CHM.
           In Eq.(\ref{external_field}), $H_0 / \left( \Phi_0 / \xi_0^2 \right)=0.01$ and $H_{{\rm appl}} / \left( \Phi_0 / \xi_0^2 \right)=0.00$.}
%          So, the amplitudes of the external magnetic field $\left( H_{{\rm ext}} \right)_z/\left( \Phi_0/\xi_0^2 \right)(x)$ is $0.01$.}
  \label{Fig5}
\end{figure}

First, we show effects of the CHM on distributions of the order parameter without the external applied magnetic field ($H_{\rm appl} = 0$).
%We take the system size as $7.0L'\xi_0\times20\xi_0$.
%In Eq.(\ref{external_field}), $H_0/(\Phi_0/\xi_0^2)$ in the first term is changed and the second term is zero.
For $H_0 / \left( \Phi_0 / \xi_0^2 \right) = 0.01$, results are shown in Fig.\ref{Fig5}.
From these results, the order parameter oscillates spatially,
%In Fig.\ref{OP}(c), $(\bm{H}_{{\rm ext}}_z)/($\Phi_0/\xi_0$)=0.01,$
although the order parameter is uniform under the weak homogeneous magnetic field.
The period of the order parameter is a half of that of the magnetic field $(\bm{H}_{\rm ext})_z$.
%Then, vortices appear in the large magnetic field region. In the vortex core, the order parameter of superconductivity is zero.
%So, we can predict that vortices appear in the top and bottom of the oscillating magnetic filed.
%Next, we show vortex states under the magnetic field from the CHM.

\subsection{Effects of the CHM on vortex states}
\begin{figure}[t]
 \centering
  \includegraphics[scale=0.25]{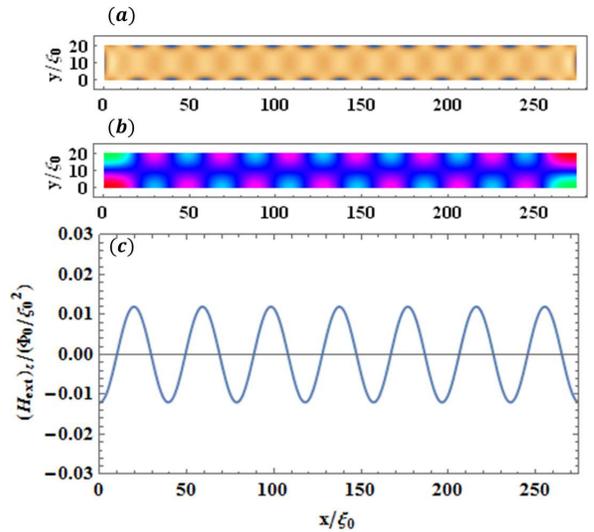}
   \caption{(a) Distributions of order parameter, (b) phase, and (c) magnetic fields.
            Magnetic fields from the chiral helimaget $H_0/(\Phi_0/\xi_0^2) = 0.012$ and the applied magnetic fields $H_{{\rm appl}}/(\Phi_0/\xi_0^2)=0.000$.}
  \label{Fig6}
\end{figure}

\begin{figure}[t]
 \centering
  \includegraphics[scale=0.25]{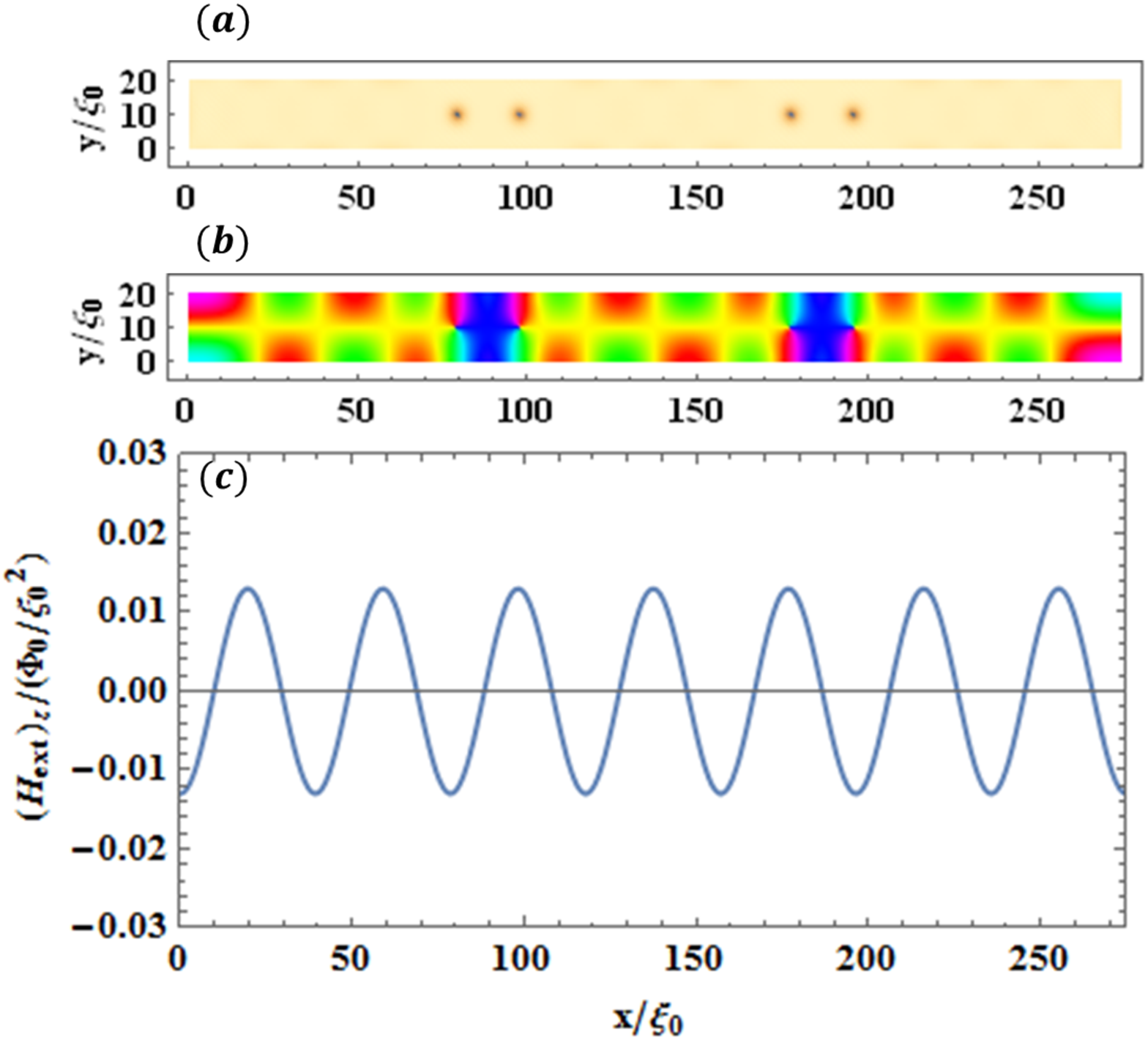}
   \caption{(a) Distributions of order parameter, (b) phase, and (c) magnetic fields.
            Magnetic fields from the chiral helimaget $H_0/(\Phi_0/\xi_0^2) = 0.013$ and the applied magnetic fields $H_{{\rm appl}}/(\Phi_0/\xi_0^2)=0.000$.}
  \label{Fig7}
\end{figure}

\begin{figure}[t]
 \centering
  \includegraphics[scale=0.25]{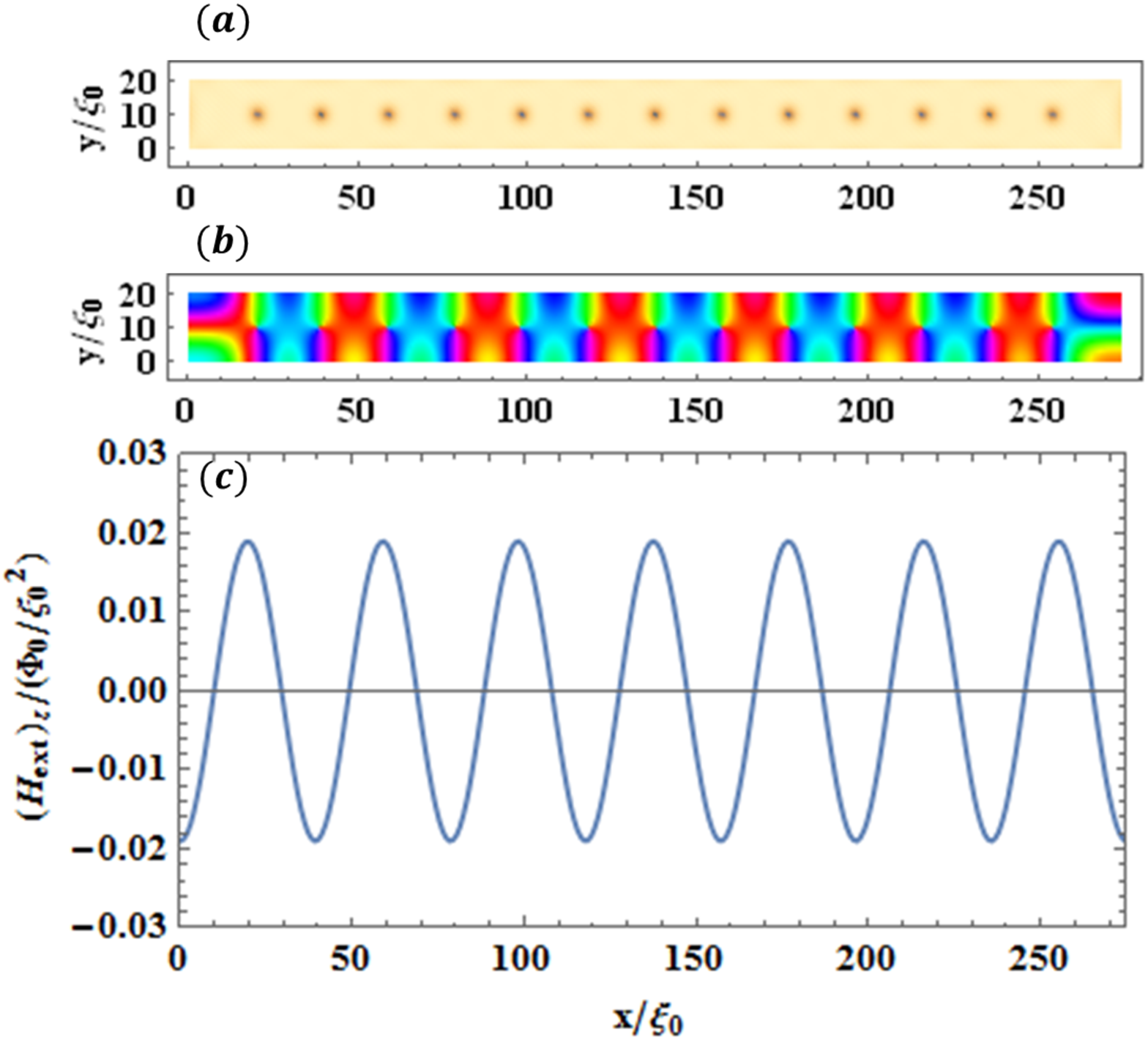}
   \caption{(a) Distributions of order parameter, (b) phase, and (c) magnetic fields.
            Magnetic fields from the chiral helimaget $H_0/(\Phi_0/\xi_0^2) = 0.019$ and the applied magnetic fields $H_{{\rm appl}}/(\Phi_0/\xi_0^2)=0.000$.}
  \label{Fig8}
\end{figure}

\begin{figure}[t]
 \centering
  \includegraphics[scale=0.25]{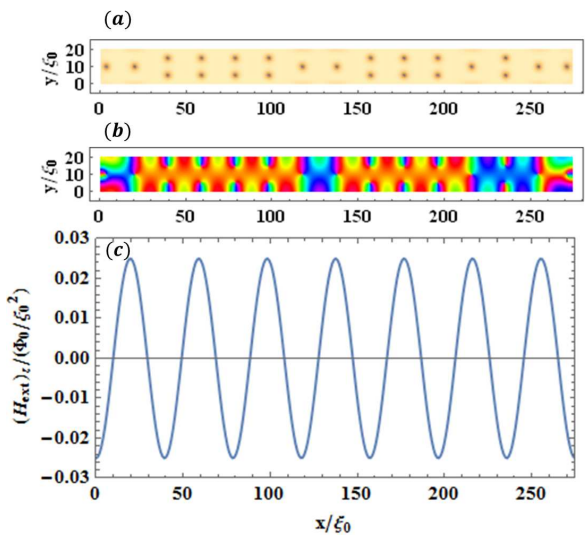}
   \caption{(a) Distributions of order parameter, (b) phase, and (c) magnetic fields.
            Magnetic fields from the chiral helimaget $H_0/(\Phi_0/\xi_0^2) = 0.025$ and the applied magnetic fields $H_{{\rm appl}}/(\Phi_0/\xi_0^2)=0.000$.}
  \label{Fig9}
\end{figure}

Next, we show effects of the CHM on vortex states without the external applied magnetic field ($H_{\rm appl} = 0$).
%In Eq.(\ref{external_field}), $H_0/(\Phi_0/\xi_0^2)$ in the first term is changed and the second term is zero.
We show vortex configurations for $H_0/(\Phi_0/\xi_0^2)=0.012$ (Fig.\ref{Fig6}), $0.013$ (Fig.\ref{Fig7}), $0.019$ (Fig.\ref{Fig8}), and $0.025$ (Fig.\ref{Fig9}).
%\begin{figure}[t]
% \centering
%  \includegraphics[scale=0.14]{2-1-0.eps}
%  \includegraphics[scale=0.14]{2-1-1.eps}
%    \caption{Distributions of order parameters [(a), (d), (g), (j)], phases [(b), (e), (h), (k)], and magnetic fields [(c), (f), (i), (l)].
%            Magnetic fields from the chiral helimaget $H_0/(\Phi_0/\xi_0^2)$ are $0.012$(I), $0.013$(II), $0.019$(III), and $0.025$(IV) and the applied magnetic fields $H_{{\rm appl}}/(\Phi_0/\xi_0^2)$ are zero.}
%  \label{2-1}
%\end{figure}

When $H_0/(\Phi_0/\xi_0^2) = 0.010$ (Fig.\ref{Fig5}) and $0.012$ (Fig.\ref{Fig6}), vortices don't appear.
However, $H_0/(\Phi_0/\xi_0^2) = 0.013$ (Fig.\ref{Fig7}), four vortices appear.
From Fig.\ref{Fig7}(b) and (c), we find two kinds of vortices whose directions of quantum fluxes are parallel to directions of magnetic fields.
In this paper, we call these vortices up- ($B_z>0$) and down-vortices($B_z<0$), respectively.
Here, $B_z$ is a magnetic flux density.
In Fig.\ref{Fig7}(a), up- and down-vortices appear next to each other.
They do not appear separately.
In addition, the pair annihilation of up- and down-vortices doesn't occur despite of the short distance between up- and down-vortices.
%We explain this question by 
Considering two interactions for vortices, this behavior can be explained.
One of these interactions is an attractive interaction between up- and down-vortices.
Due to this attractive interaction, these vortices tend to approach each other.
Another is an interaction between the vortex and the magnetic field.
Due to this interaction, the vortex tends to appear in the large magnetic field region.
From the competition of these two interactions, up- and down-vortices approach each other, but remain in the stronger field region.
Therefore, the pair annihilation doesn't occur and up- and down-vortex appear next to each other.

For larger field from the CHM (Fig.\ref{Fig8} and \ref{Fig9}), the number of vortices increases.
From the distributions of phases and magnetic field, up- and down-vortices appear alternately.
These configurations can be explained by the same discussion about interactions of vortices.
Generally, up- and down-vortices appear in the parallel magnetic field region.

\begin{figure}[htbp]
 \centering
  \includegraphics[scale=0.25]{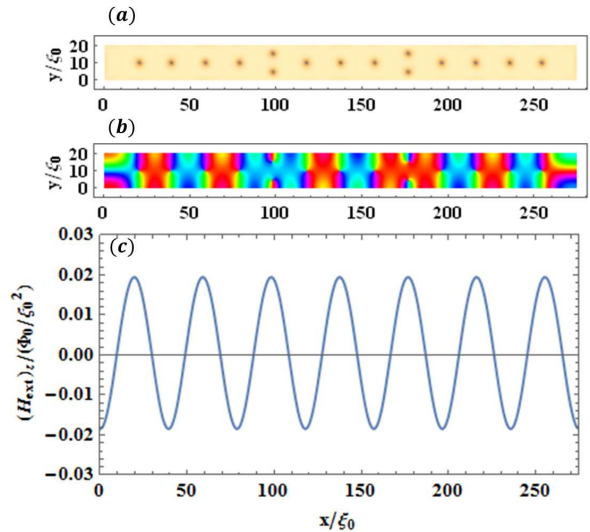}
   \caption{(a) Distributions of order parameter, (b) phase, and (c) magnetic fields.
            Magnetic fields from the chiral helimaget $H_0/(\Phi_0/\xi_0^2) = 0.019$ and the applied magnetic fields $H_{{\rm appl}}/(\Phi_0/\xi_0^2)=0.0005$.}
  \label{Fig10}
\end{figure}

\begin{figure}[htbp]
 \centering
  \includegraphics[scale=0.25]{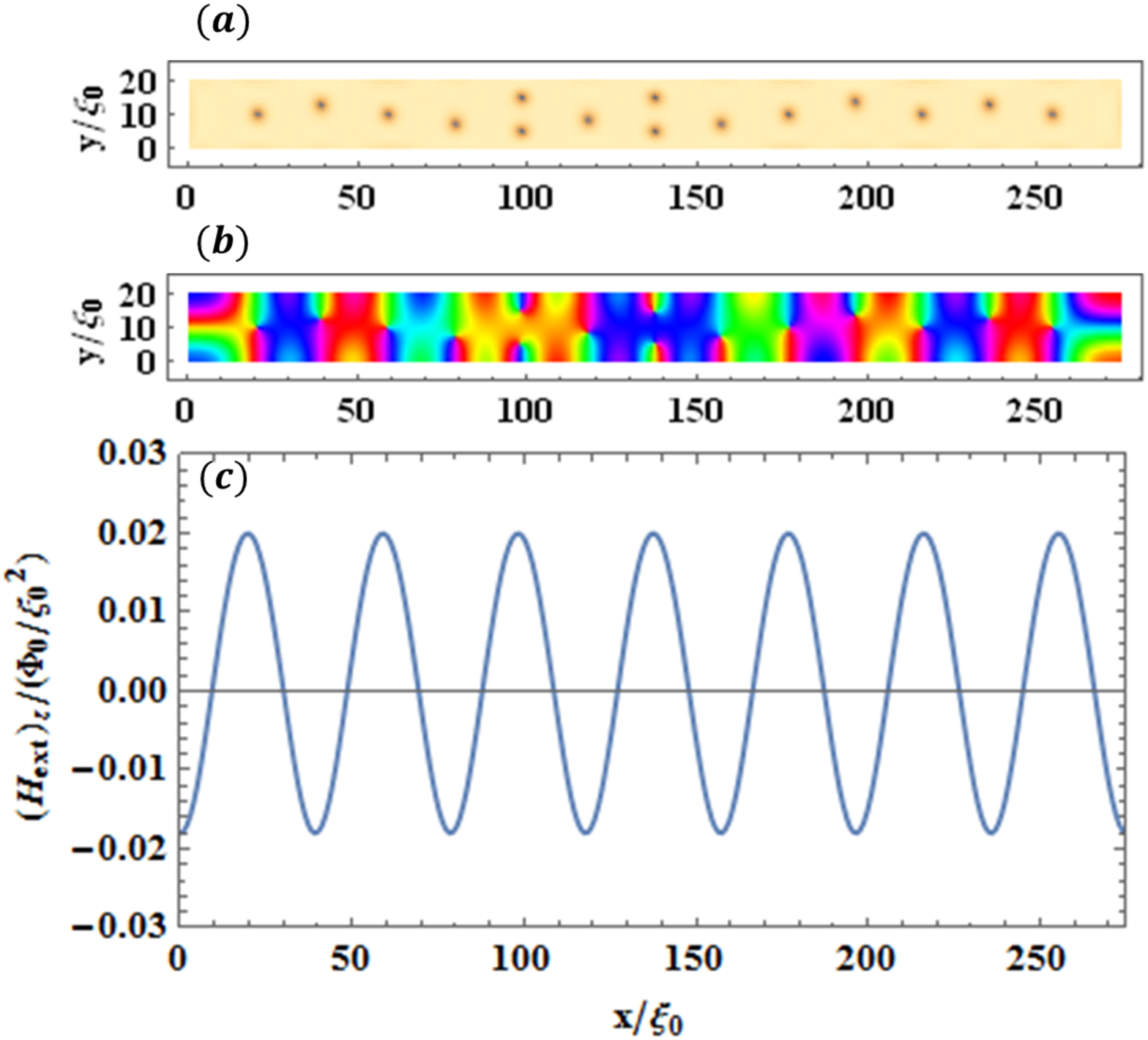}
   \caption{(a) Distributions of order parameter, (b) phase, and (c) magnetic fields.
            Magnetic fields from the chiral helimaget $H_0/(\Phi_0/\xi_0^2) = 0.019$ and the applied magnetic fields $H_{{\rm appl}}/(\Phi_0/\xi_0^2)=0.0010$.}
  \label{Fig11}
\end{figure}

\begin{figure}[htbp]
 \centering
  \includegraphics[scale=0.25]{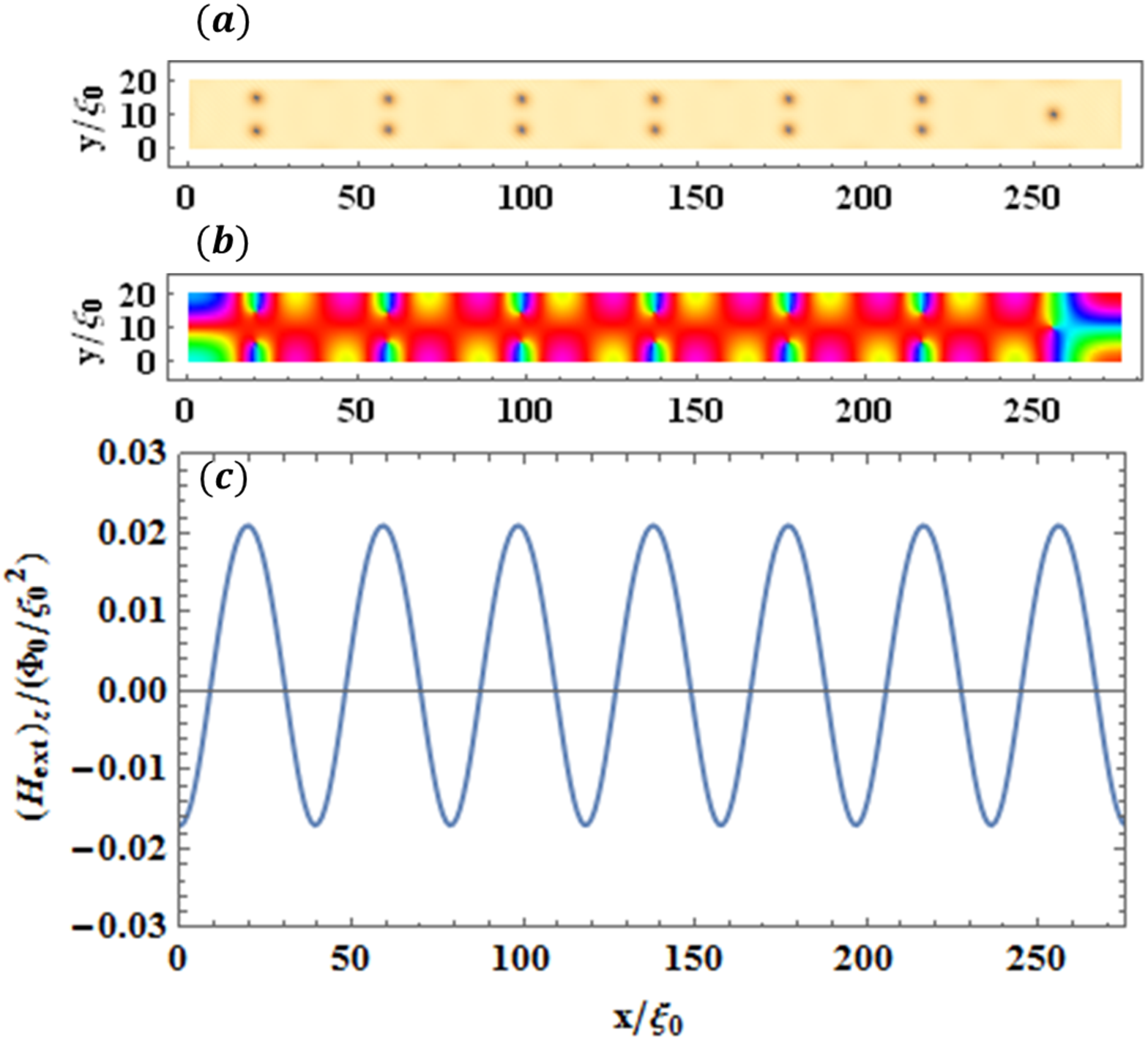}
   \caption{(a) Distributions of order parameter, (b) phase, and (c) magnetic fields.
            Magnetic fields from the chiral helimaget $H_0/(\Phi_0/\xi_0^2) = 0.019$ and the applied magnetic fields $H_{{\rm appl}}/(\Phi_0/\xi_0^2)=0.0020$.}
  \label{Fig12}
\end{figure}

\begin{figure}[htbp]
 \centering
  \includegraphics[scale=0.25]{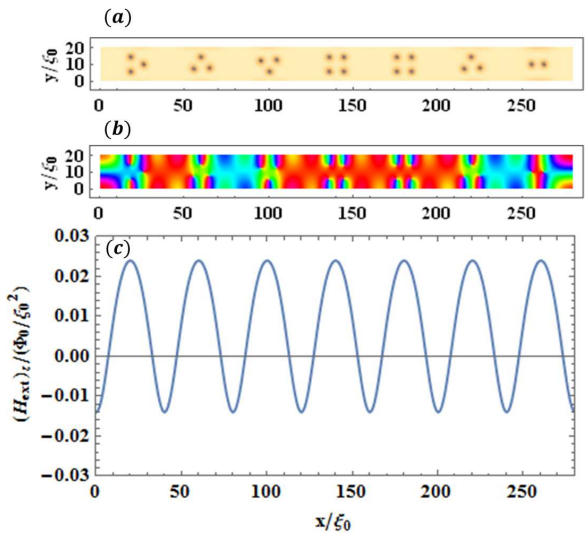}
   \caption{(a) Distributions of order parameter, (b) phase, and (c) magnetic fields.
            Magnetic fields from the chiral helimaget $H_0/(\Phi_0/\xi_0^2) = 0.019$ and the applied magnetic fields $H_{{\rm appl}}/(\Phi_0/\xi_0^2)=0.0050$.}
  \label{Fig13}
\end{figure}

\subsection{Effects of the CHM and the applied magnetic field on vortex states}
So far, we have discussed vortex states under the magnetic field from the CHM without homogeneous applied magnetic field.
Next, we show effects of the CHM and the applied magnetic field on vortex states.
In the following, $H_0/(\Phi_0/\xi_0^2)$ is fixed to $0.019$.
The vortex state under the zero applied magnetic field has been already shown in Fig.\ref{Fig8}.
We show vortex configurations for $H_{{\rm appl}}/(\Phi_0/\xi_0^2) = 0.0005$ (Fig.\ref{Fig10}), $0.001$ (Fig.\ref{Fig11}), $0.002$ (Fig.\ref{Fig12}) and $0.005$ (Fig.\ref{Fig13}).
%\begin{figure}[t]
% \centering
%  \includegraphics[scale=0.14]{3-1-0.eps}
%  \includegraphics[scale=0.14]{3-1-1.eps}
%    \caption{Distributions of order parameters [(a), (d), (g), (j)], phases [(b), (e), (h), (k)], and magnetic fields [(c), (f), (i), (l)].
%            Magnetic fields from the chiral helimaget $H_0/(\Phi_0/\xi_0^2)$ are fixed to $0.019$ and the applied magnetic fields $H_{{\rm appl}}/(\Phi_0/\xi_0^2)$ are $0.0005$ (I), $0.001$ (II), $0.002$ (III), and $0.005$ (IV).}
%  \label{2-2}
%\end{figure}

In Fig.\ref{Fig10}, the number of up-vortices increases due to the applied magnetic field.
For $H_{{\rm appl}}/(\Phi_0/\xi_0^2) = 0.001$ (Fig.\ref{Fig11}), down-vortices approach to edges.
When the applied magnetic field increases, for $H_{{\rm appl}}/(\Phi_0/\xi_0^2)=0.002$ (Fig.\ref{Fig12}) and $0.005$ (Fig.\ref{Fig13}) down-vortices disappear.
In Fig.\ref{Fig12}, the total external magnetic field $H_{{\rm ext}}/(\Phi_0/\xi_0^2)$ oscillates between $-0.017$ and $0.021$.
Because the absolute value of the negative magnetic field becomes small ($|-0.017|$), the interaction between a down-vortex and the external magnetic field becomes weaker than that between up- and down-vortices.
So, down-vortices gradually disappear with increasing the applied magnetic field.

For larger applied magnetic field, the number of up-vortices increases, 
and they form parallel, triangular, or square structures.
These structures of up-vortices are stable in the positive magnetic field region, and they don't prefer to appear in the negative magnetic field region. 
If we apply an external current along $y$-axis, up-vortices can't move through the negative magnetic field region.
Therefore, we expect the pinning effect of the vortex due to the helical magnetic structure, which leads to the increase of the critical current.

\section{Summary \label{Summary}}
We have investigated vortex states in the CHM / SC bilayer systems using two-dimensional Ginzburg-Landau equations.
We found that up-vortices and down-vortices appear alternately under the magnetic field from the CHM.
Moreover, when the homogeneous magnetic field ($>0$) is applied, down-vortices disappear and the number of vortices increases in the positive magnetic field region.
Then, up-vortices form parallel, triangular, or square structures.
The pinning effect on vortices is expected from these configurations, which leads to the increase of the critical current.
So, investigating the dynamics of vortices in the CHM / SC bilayer system solving time-dependence Ginzburg-Landau equations is a future problem.

In this study, we have solved two-dimensional Ginzburg-Landau equations.
Then, the effect of the CHM has been taken as only the $z$-component of the magnetic field.
So, we don't still treat the magnetic structure of the CHM completely.
In the future, we will solve GL equations for the three-dimensional bilayer system in order to investigate the effect of the chirality of the CHM on the bilayer system.
\begin{acknowledgments}
This work was supported by JPSJ KAKENHI Grant Number 26400367.
\end{acknowledgments}

\appendix*
\section{}
In this appendix, we give coefficients in Eqs.(\ref{gl_3})-(\ref{gl_6}).
They are defined as,
\begin{eqnarray}
 & & P_{ij}(\{ \mbox{\boldmath $A$} \}, \{ \psi \}) \equiv K_{ij}^{xx} + K_{ij}^{yy} + \sum_{i_1, i_2} I_{i_1, i_2, i, j} \left( A_{i_1x}^e A_{i_2x}^e \right. \nonumber \\
 & & \left. + A_{i_1y}^e A_{i_2y}^e \right) - \frac{1}{\xi^2(T)} I_{ij} \\
 & & P_{ij}^{2R} (\{ \psi \}) \equiv \frac{1}{\xi^2(T)} \sum_{i_1, i_2} I_{i_1, i_2, i, j} \left( 3 {\rm Re}~\psi_{i_1}^e {\rm Re}~\psi_{i_2}^e \right. \nonumber \\
 & & \left. +  {\rm Im}~\psi_{i_1}^e {\rm Im}~\psi_{i_2}^e \right) \\
 & & P_{ij}^{2I} (\{ \psi \}) \equiv \frac{1}{\xi^2(T)} \sum_{i_1, i_2} I_{i_1, i_2, i, j} \left(   {\rm Re}~\psi_{i_1}^e {\rm Re}~\psi_{i_2}^e \right. \nonumber \\
 & & \left. + 3{\rm Im}~\psi_{i_1}^e {\rm Im}~\psi_{i_2}^e \right) \\
 & & Q(\{\mbox{\boldmath $A$}\}) \equiv \sum_{i_1} \left( J_{j, i_1, i}^x A_{i_1}^x + J_{j, i_1, i}^y A_{i_1}^y \right. \nonumber \\
 & & \left. - J_{i, i_1, j}^x A_{i_1}^x - J_{i, i_1, j}^y A_{i_1}^y \right) \\
 & & Q^2(\{\psi\}) \equiv \frac{2}{\xi^2} \sum_{i_1, i_2} I_{i_1, i_2, i, j} {\rm Im}~\psi_{i_1}^e {\rm Re}~\psi_{i_2}^e \\
 & & R_{ij} (\{\psi\}) \equiv \kappa^2\xi^2(T) \left( K_{ij}^{xx} + K_{ij}^{yy} \right)  \nonumber \\
 & & + \sum_{i_1, i_2} I_{i_1, i_2, i, j} \psi_{i_1}^{e\ast} \psi_{i_2}^e \\
 & & S_{ij} \equiv \kappa^2 \xi^2(T) \left( K_{ij}^{xy} - K_{ij}^{yx} \right) \\
 & & T_i^{x_i} \equiv \sum_{i_1, i_2} J_{i_1, i_2, i}^{x_i} {\rm Im}~\left( \psi_{i_1}^{e\ast} \psi_{i_2}^e \right) \\
 & & U_i^{x_i} \equiv \kappa^2 \xi^2(T) \frac{2\pi}{\Phi_0} H J_i^{x_i} \\
 & & V_i^R(\{\psi\}) \equiv \frac{2}{\xi^2(T)} \sum_{i_1, i_2, i_3} I_{i_1, i_2, i_3, i} \psi_{i_1}^e \psi_{i_2}^{e\ast} {\rm Re}~\psi_{i_3}^e \\
 & & V_i^I(\{\psi\}) \equiv \frac{2}{\xi^2(T)} \sum_{i_1, i_2, i_3} I_{i_1, i_2, i_3, i} \psi_{i_1}^e \psi_{i_2}^{e\ast} {\rm Im}~\psi_{i_3}^e. 
\end{eqnarray}
Their coefficients can be calculated by integrals $I_{ij}^e,I_{i_1,i_2,i_3}^e,I_{i_1,i_2,i_3,i_4}^e,J_{i_1,i_2,i_3}^{e_x},J_{i_1,i_2,i_3}^{e_y},J_j^{e_{x_i}},$ and $K_{i_1,i_2}^{e_{x_i},e_{x_j}}$.
Their integrals are given by,
\begin{eqnarray}
 & & I_{ij}^e \equiv \int_{\Omega_e} N_i^e N_j^e {\rm d}\Omega \\
 & & I_{i_1, i_2, i_3}^e \equiv \int_{\Omega_e} N_{i_1}^e N_{i_2}^e N_{i_3}^e {\rm d}\Omega \\
 & & I_{i_1, i_2, i_3, i_4}^e \equiv \int_{\Omega_e} N_{i_1}^e N_{i_2}^e N_{i_3}^e N_{i_4}^e {\rm d}\Omega \\
 & & J_{i_1, i_2, i_3}^{e_x} \equiv \int_{\Omega_e} \frac{\partial N_{i_1}^e}{\partial x} N_{i_2}^e N_{i_3}^e {\rm d}\Omega \\
 & & J_{i_1, i_2, i_3}^{e_y} \equiv \int_{\Omega_e} \frac{\partial N_{i_1}^e}{\partial y} N_{i_2}^e N_{i_3}^e {\rm d}\Omega \\
 & & J_j^{e_{x_i}} \equiv \int_{\Omega_e} \frac{\partial N_j^e}{\partial x_i}{\rm d}\Omega \\
 & & K_{i_1, i_2}^{e_{x_i, x_j}} \equiv \int_{\Omega_e} \frac{\partial N_{i_1}^e}{\partial x_i} \frac{\partial N_{i_2}^e}{\partial x_j} {\rm d} \Omega.
\end{eqnarray}

\end{document}